\documentclass[floatfix,prd,showpacs,showkeys,preprintnumbers,nofootinbib,superscriptaddress,11pt]{revtex4-1}
\usepackage[utf8]{inputenc}
\usepackage[sort&compress]{natbib}
\usepackage{ulem}
\usepackage{bm}
\usepackage{times}
\usepackage{amssymb,amsbsy,amsmath,amsfonts}
\usepackage{graphicx}
\usepackage{epstopdf}
\usepackage{float}
\usepackage{color}
\usepackage{morefloats}
\usepackage{rotating}
\usepackage{srcltx}
\usepackage{slashed}
\usepackage{subfigure}
\usepackage{multirow}
\usepackage{verbatim}
\usepackage{hyperref}
\usepackage{overpic}
\usepackage{tabularx}
\renewcommand{\arraystretch}{1.2}

\begin{document}

\title{ Decay constants of the two-pole $D_0^*(2300)$}

\author{Qi-Wei Yuan}
\affiliation{School of Nuclear Science and Technology, Lanzhou University, Lanzhou 730000, China}

\author{Jia-Ting Zhang}
\affiliation{School of Nuclear Science and Technology, Lanzhou University, Lanzhou 730000, China}

\author{Ming-Zhu Liu}~\email{liumz@lzu.edu.cn}
\affiliation{Frontiers Science Center for Rare Isotopes, Lanzhou University, Lanzhou 730000, China}
\affiliation{School of Nuclear Science and Technology, Lanzhou University, Lanzhou 730000, China}

\begin{abstract}
The nature of the scalar charmed meson $D_0^*(2300)$ remains one of the most intriguing questions in hadron spectroscopy. A prominent interpretation is a two-pole structure generated by the coupled channels $D\pi$, $D\eta$, $D_s\bar K$, and $D\eta'$, in which the lower pole couples mainly to $D\pi$ and the higher pole to $D_s\bar K$. Following this picture, we calculate the decay constants of these two poles using the effective Lagrangian approach. The resulting values, 
$f_{D_0^*}^{(\mathrm{lower})}=68.46^{+0.93}_{-1.25}\,\mathrm{MeV}$ and
$f_{D_0^*}^{(\mathrm{higher})}=90.51^{+5.91}_{-11.64}\,\mathrm{MeV}$, are substantially smaller than those predicted by conventional $c\bar q$ excited-state scenarios. This difference suggests that decay constants can serve as a sensitive probe to discriminate  between different internal structures of the $D_0^*(2300)$. As a phenomenological application, we further predict the branching fractions for the Cabibbo-favored decays $B_s \to D_s D_0^*$, $\Lambda_b \to \Lambda_c D_0^*$, and $\Xi_b \to \Xi_c D_0^*$ within the factorization approach. These predictions provide crucial tests to validate the two-pole interpretation of the $D_0^*(2300)$.    
\end{abstract}


\maketitle

\section{Introduction}

A growing number of hadronic states beyond conventional mesons and baryons, i.e., exotic states,  have been observed in experiments~\cite{Brambilla:2010cs,Olsen:2017bmm,Brambilla:2019esw}, challenging our understanding of hadronic structure but offering a chance to probe nonperturbative aspects of the strong interaction~\cite{Chen:2016qju,Lebed:2016hpi,Oset:2016lyh,Esposito:2016noz,Dong:2017gaw,Guo:2017jvc,Ali:2017jda,Karliner:2017qhf,Guo:2019twa,Meng:2022ozq,Liu:2024uxn,Wang:2025sic,Doring:2025sgb}. Studies of    exotic-state properties indicate that hadron-hadron interaction, as a fundamental nonperturbative effect in quantum chromodynamics, plays a significant role  both in the dynamical formation of exotic hadronic structures and in the quantitative understanding of the production rates of exotic states in  weak decays~\cite{Liu:2006df,Wu:2019rog,Wu:2023rrp,Wu:2025fzx}.  Recent investigations have further demonstrated that coupled-channel effects and the dressing effect of bare QCD states can substantially modify hadron-hadron interactions, thereby leading to complex hadronic structures \cite{Bai:2026atm}.

The two-pole structure is an interesting hadronic phenomenon that is dynamically generated by coupled-channel potentials. A typical example is $\Lambda(1405)$~\cite{Alston:1961zzd}. The studies based on Lattice QCD simulation and covariant chiral perturbation theory   conclude that there exist two states in the vicinity of  $\Lambda(1405)$~\cite{Lu:2022hwm,BaryonScatteringBaSc:2023zvt}, where the higher pole couples dominantly  to the $\bar{K}N$ channel, and the lower pole couples dominantly to the $\pi\Sigma$ channel.  Another typical example is $D_0^*(2300)$~\cite{Meissner:2020khl}.  This two-pole structure is generated by  the coupled channels $D\pi$, $D\eta$, $D_s\bar{K}$, and $D\eta'$ , where the lower pole is mainly associated with the  $D\pi$ channel, whereas the higher pole is closely related to the $D_s\bar{K}$ channel~\cite{vanBeveren:2003kd,Kolomeitsev:2003ac,Gamermann:2006nm,Guo:2006fu,Guo:2009ct,Guo:2015dha,Du:2017zvv,Guo:2018tjx}.

The underlying mechanisms for the two-pole structures of the $\Lambda(1405)$ and $D_0^*(2300)$ are rooted in the complex dynamics of strong interactions. Chiral dynamics plays an essential role in their generation. As indicated in Ref.~\cite{Xie:2023cej}, the energy-dependent terms constrained by chiral dynamics, i.e., Weinberg-Tomozawa interactions,   are specifically responsible for generating these two-pole structures. In addition, the coupled-channel potentials  constrained by  SU(3)-flavor symmetry  play the dominant role in generating  two-pole structures~\cite{Xie:2023cej,Zhuang:2024udv}. Under SU(3)-flavor symmetry,  the lower pole and higher pole for $D_0^*(2300)$ originate from the group representations of $\bar{3}$ and $6$, respectively~\cite{Albaladejo:2016lbb},  whereas those for   $\Lambda(1405)$ arise from distinct group representations of $1$ and $8$, respectively~\cite{Jido:2003cb,Guo:2023wes}.  We emphasize that this assignment refers to the dynamical origin of the poles
in the SU(3)-flavor-symmetric limit and to their adiabatic connection to it~\cite{Albaladejo:2016lbb}. At the physical point, explicit SU(3) breaking renders each pole's wave
function a mixture of both irreps.   Therefore,   the emergence of two-pole structures is very common in  
coupled-channel chiral dynamics~\cite{Meissner:2020khl}.  



In this work, we only focus   on the two-pole structure $D_0^*(2300)$.  This structure persists in the analysis of lattice QCD simulation data from Refs.~\cite{Moir:2016srx,Yan:2024yuq,Luo:2026kui,Zhuang:2026lta}. Consequently, investigating other physical observables associated with this state is crucial for corroborating its two-pole nature.  Du et al.~have previously verified this structure via an analysis of angular moments in 
$b$-flavored meson decays~\cite{Du:2017zvv,Du:2019oki}, and have recently proposed to determine it from semileptonic decays~\cite{Du:2025beb}. Lyu et al. suggested that the two-pole structure produces a significant signal in the invariant mass distribution of 
$B$-meson decays~\cite{Lyu:2023ppb}.   In addition, femtoscopy
  is a developing method for studying the interactions between hadrons~\cite{Fabbietti:2020bfg,ALICE:2020mfd}, and was proposed to study the two-pole structure $D_0^*(2300)$ by measuring the corresponding momentum correlation functions~\cite{ALICE:2024bhk,Torres-Rincon:2023qll}.     It is therefore important to explore additional physical observables of  $D_0^*(2300)$. 

The decay constant of a hadron is a key quantity for characterizing its production in vacuum. Conventionally, the decay constant of the exotic meson $D_{s0}^*(2317)$ is extracted from its two-point correlation function under the assumption that it is a conventional excited-state hadron~\cite{Verma:2011yw,Wang:2015mxa,Wang:2007av}. Alternatively, Refs.~\cite{Faessler:2007cu,Liu:2023cwk} developed an approach to compute this decay constant by treating the $D_{s0}^*(2317)$ as a hadronic molecule. Consequently, the decay constant is expected to be sensitive to the internal structure of the $D_{s0}^*(2317)$.
Following Refs.~\cite{Faessler:2007cu,Liu:2023cwk}, we adopt the effective Lagrangian approach to compute the decay constant of the excited charmed meson $D_0^*(2300)$, which is assumed to be a two-pole structure. To further provide relevant physical observables to be measured by experiments, we evaluate their production rates in $b$-flavored hadron decays, considering only the Cabibbo-favored modes, which can be reliably studied within the naive factorization approach~\cite{Chau:1982da,Bauer:1986bm}.



This work is organized as follows. 
In Sec.~\ref{II}, we present the theoretical framework. 
We first describe the calculation of the decay constant of  the $D_0^*(2300)$ within the effective Lagrangian approach, assuming it exhibits a two-pole structure. We then introduce the unitarized amplitude derived from chiral effective field theory~(ChEFT), which generates the two-pole structure and enables the extraction of the corresponding pole residues. Finally, we present the factorization formulas for the Cabibbo-favored decays. In Sec.~\ref{III}, we provide the numerical results, including the pole positions, molecular couplings to their constituents, scattering lengths, decay constants, and branching fractions. A brief summary is given in Sec.~\ref{IV}.

\section{Theoretical framework}\label{II}

The primary goal of this work is to calculate the decay constant of the excited charmed meson $D_0^*(2300)$. This state is assumed to have a two-pole structure generated by the unitarized coupled-channel amplitude for $D\pi$, $D\eta$, $D_s\bar{K}$, and $D\eta'$, with the interactions derived from ChEFT and iterated through the Lippmann-Schwinger~(LS) equation. We employ the effective Lagrangian approach to compute the decay constants of the two-pole $D_0^*(2300)$.  As phenomenological applications, we evaluate the branching fractions of $B_s$, $\Lambda_b$, and $\Xi_b$ decays into $D_0^*(2300)$ accompanied by other hadrons, using the calculated decay constants.

\subsection{Decay constants in the molecular picture}\label{sec:decayconstants}

\begin{figure}[ttt]
\centering
\includegraphics[width=0.50\textwidth]{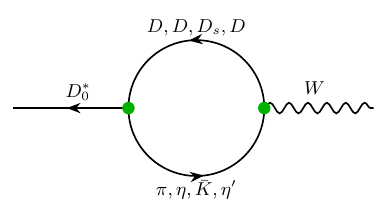}
  \caption{Feynman diagram for the transition of the $W$ boson into $D_0^*$ in the $D\pi$, $D\eta$, $D_s\bar{K}$, and $D\eta'$ molecular picture.}
  \label{W}
\end{figure}

For the scalar charmed state, its decay constant is defined by the following vector-current matrix element
\begin{equation}\label{eq:decayconstant}
\langle D_{0}^{*}(p)|J^{\mu}|0\rangle=f_{D_{0}^{*}}p^{\mu},
\end{equation}
where $f_{D_{0}^{*}}$ and $p^{\mu}$ represent the decay constant and momentum of $D_{0}^{*}(2300)$, respectively.    { In the molecular picture adopted here, where the $D_{0}^{*}(2300)$ emerges dynamically as a two-pole structure from coupled-channel hadron--hadron interactions~\cite{Guo:2015dha}, the effective current $J^{\mu}=(\bar{q}\Gamma q)\,\gamma^{\mu}\,(\bar{q}\Gamma' c)$ probed by the higher Fock components of the state is more naturally identified with a composite operator reflecting the underlying two-hadron dynamics, where $\Gamma$ and $\Gamma'$ denote appropriate Dirac matrices. The light-quark bilinear $\bar{q}\Gamma q$ transforms as an octet ($\mathbf{8}$) (denoted by $P$), while the charm-containing bilinear $\bar{q}\Gamma' c$ transforms as an anti-triplet ($\bar{\mathbf{3}}$) (denoted by $D_{(s)}$). Their tensor product,
\begin{equation}\label{eq:decomp}
\mathbf{8}\otimes\bar{\mathbf{3}}=\bar{\mathbf{3}}\oplus\mathbf{6}\oplus\overline{\mathbf{15}},
\end{equation}
necessarily contains both $\bar{\mathbf{3}}$ and $\mathbf{6}$ irreducible components.  Consequently, the composite current can generate both the $\bar{\mathbf 3}$
and the $\mathbf 6$ components of the pole wave functions. In contrast, the
local current $(c \bar q)$ entering the factorization formulas of
Sec.~\ref{sec:weakapplications} is a pure $\bar{\mathbf 3}$ operator: its matrix element
with a pole probes only the $\bar{\mathbf 3}$ component of that pole's wave
function and vanishes for a pure $\mathbf 6$ state in the SU(3)-flavor
symmetric limit~\cite{Albaladejo:2016lbb,Zhuang:2026lta}.   The matrix element in Eq.~(\ref{eq:decayconstant}) is illustrated by a loop diagram in Fig.~\ref{W}, where the left vertex corresponds to the generation of the two-pole $D_{0}^*(2300)$ through the interactions between $D_{(s)}$ and $P$ mesons ($P=\pi,\eta,\bar K,\eta'$), whereas the right vertex describes the transition of the $D_{(s)}$ mesons to the $P$ mesons and  $W$ boson.   In general, the decay constant $f_{D_{0}^{*}}$ is extracted from the coefficient of $p^{\mu}$ in the amplitude obtained from Fig.~\ref{W}.  In this work, we employ the effective Lagrangian approach to compute the  corresponding amplitude. }  


The effective Lagrangian for the right vertex of Fig.~\ref{W}, which describes the  $D_{(s)}$ mesons  transition to the $P$ meson and $W$-boson, is written as~\cite{Liu:2023cwk}
\begin{eqnarray}
\mathcal{L}_{VD_{(s)}P}&=f^{D_{(s)}P}(q^2)V^{\mu}(D_{(s)}\partial_{\mu}P-\partial_{\mu}D_{(s)}P),
\end{eqnarray}
where $V^\mu$ denotes the external weak vector current and
$f^{D_{(s)}P}(q^2)$ the $D_{(s)}\to P$ transition form factor induced
by the weak current.   Following Ref.~\cite{Faessler:2007cu}, we approximately  take
the form factor at zero momentum transfer, $f^{D_{(s)}P}(0)$.  The form factors $f^{D\pi}(0)$, $f^{D\eta}(0)$,
$f^{D_s\bar K}(0)$, and $f^{D\eta'}(0)$ are determined from the
corresponding semileptonic branching fractions. Using the experimental
data of Ref.~\cite{ParticleDataGroup:2026mpi}, we obtain
$f^{D\pi}(0)=0.706\pm0.005$, $f^{D\eta}(0)=0.394\pm0.012$,
$f^{D_s\bar K}(0)=0.709^{+0.029}_{-0.033}$, and
$f^{D\eta'}(0)=0.324\pm0.016$.

The effective Lagrangian for the left vertex in Fig.~\ref{W}, which describes the formation of the two-pole $D_0^*(2300)$ from the interactions between $D_{(s)}$ and $P$ mesons, is given by~\cite{Liu:2023cwk}
\begin{eqnarray}
\mathcal{L}_{D_{0}^*D_{(s)}P}&=g_{D_{0}^*D_{(s)}P}D_{0}^*D_{(s)}P,
\end{eqnarray}
where $g_{D_{0}^*D\pi}$, $g_{D_{0}^*D\eta}$, $g_{D_{0}^*D_s\bar{K}}$, and $g_{D_{0}^*D\eta'}$ denote the couplings of the $D_0^*$ to its constituent mesons. These couplings are determined from the pole residues of the unitarized amplitude~\cite{Guo:2008gp}, which is widely used in studies of hadronic molecules~\cite{Xiao:2019aya,Du:2019pij,Pan:2023hrk}.

With the above Lagrangians, the amplitude represented by Fig.~\ref{W} is written as  
\begin{equation}
\label{ds2317int}
\mathcal{A}=g_{D_{0}^*D_{(s)}P} f^{D_{(s)}P}(0)\int \frac{d^4 k_1}{(2\pi)^4}\frac{1}{k_{1}^2-m_{1}^2}\frac{1}{k_{2}^2-m_{2}^2}(k_{1}^{\mu}-k_{2}^{\mu}),
\end{equation}
where the subscripts 1 and 2 denote the $D_{(s)}$ and $\pi(\eta,\bar{K},\eta')$ mesons, respectively.

Once the amplitudes in Fig.~\ref{W} are obtained, the decay constants $f_{D_{0}^*}$ can be extracted from their definition.  Since the above two-point loop function is divergent,  we adopt the dimensional regularization approach to deal with the divergence. Using Feynman parameters, the above integral is obtained as 
\begin{equation}
\label{decayconstantv}
\int \frac{d^4k_1}{(2\pi)^4} \frac{k_1^{\mu}-k_2^{\mu}}{(k_1^2-m_1^2)[(p-k_1)^2-m_2^2]}=\frac{p^{\mu}}{16\pi^2}\int_{0}^{1}dx (2x-1) \ln\frac{\Delta^2}{\mu^2}, 
\end{equation}
where $p=k_1+k_2$ and $\mu$ is the renormalization scale. { We define
\begin{equation}
\Delta^2(x;p^2)
=p^2x^2-m_1^2(x-1)-x(p^2-m_2^2).
\end{equation}}
According to the definition in Eq.~(\ref{eq:decayconstant}),  the decay constant is therefore written as
\begin{equation}
\label{fd2317coupling}
f_{D_{0}^*}^{m_1 m_2}=g_{D_{0}^* m_1 m_2} f^{m_1 m_2}(0)\frac{1}{16\pi^2}\int_{0}^{1}dx (2x-1) \ln\frac{\Delta^2}{\mu^2}.   
\end{equation} 
For each pole of $D_{0}^*$, its decay constant  is obtained by summing the contributions of each component $D\pi$, $D\eta$, $D_s\bar{K}$, and $D\eta'$.  The flavor structure of the residue couplings $\boldsymbol g^{(p)}$
entering this sum is analyzed in terms of effective
antitriplet--sextet mixing angles shown later.

\subsection{Unitarized coupled-channel amplitude}\label{sec:polecouplings}

In this subsection, we  briefly introduce how to determine the  coupling between a hadronic molecule and its constituents via the residue of a scattering amplitude~\cite{Gamermann:2006nm,Guo:2009ct,Guo:2015dha}. 
The unitarized  coupled-channel scattering amplitude $T$ is obtained by solving the LS equation
\begin{eqnarray}
T(s)=(1-V(s)G(s))^{-1}V(s),
\label{lsequation}
\end{eqnarray}
where \(G(s)\) is the loop function of the two-body propagator. We evaluate
the loop function in the dimensional regularization scheme, following the same
regularization convention in Eq.~(\ref{decayconstantv}). The loop integral in the dimensional regularization scheme is given by  
\begin{eqnarray}
\label{loopfunctionv}
G(s)^{D_{0}^*} &=&\frac{1}{16\pi^2}\int_{0}^{1}dx  \ln\frac{\Delta^2}{\mu^2}.
\end{eqnarray}
{ In the two-meson scattering loop, the corresponding quantity $\Delta^2(x;s)$ is obtained from the same definition by the replacement $p^2\to s$.}
Here, $V$ denotes the potential for the scattering process $D_{(s)1}(p_1)+P_1(p_2) \to D_{(s)2}(p_3)+P_2(p_4)$ up to next-to-leading order (NLO) in ChEFT~\cite{Guo:2015dha}:
\begin{equation}
V(s, t, u)=\frac{1}{F^{2}}[\frac{C_{\text{LO}}}{4}(s-u)-4 C_{0} h_{0}+2 C_{1} h_{1}-2 C_{24} H_{24}(s, t, u)+2 C_{35} H_{35}(s, t, u)],
\end{equation} 
where $s$, $t$, and $u$ are the standard Mandelstam variables,   \(F\) is the pseudo-Nambu-Goldstone-boson decay constant,  and the  momentum-dependent functions are
\begin{equation}
\begin{aligned}
H_{24}(s,t,u)&=2h_2 p_2\cdot p_4+h_4(p_1\cdot p_2\,p_3\cdot p_4+p_1\cdot p_4\,p_2\cdot p_3),\\
H_{35}(s,t,u)&=h_3 p_2\cdot p_4+h_5(p_1\cdot p_2\,p_3\cdot p_4+p_1\cdot p_4\,p_2\cdot p_3).
\end{aligned}
\end{equation}
$h_i$ are the unknown   low-energy constants at NLO in ChEFT. 


The parameters  $C_{\rm LO}$, $C_0$, $C_1$, $C_{24}$, and $C_{35}$ are constrained by SU(3)-flavor symmetry and Chiral symmetry, whose values are collected in Table~\ref{tab:potential-coefficients}. The parameters in Table~\ref{tab:potential-coefficients} are defined as 
\begin{equation}
\begin{aligned}
X_\eta&=\sqrt{2}s_\theta-c_\theta,\\
X_{\eta'}&=\sqrt{2}c_\theta+s_\theta,\\
Z_{\eta\eta'}&=\sqrt{2}c_\theta^2-c_\theta s_\theta-\sqrt{2}s_\theta^2,\\
A_\eta&=\frac{c_\theta^2(4M_K^2-M_\pi^2)+4\sqrt{2}c_\theta s_\theta(M_K^2-M_\pi^2)+s_\theta^2(2M_K^2+M_\pi^2)}{3},\\
A_{\eta'}&=\frac{s_\theta^2(4M_K^2-M_\pi^2)+4\sqrt{2}c_\theta s_\theta(M_\pi^2-M_K^2)+c_\theta^2(2M_K^2+M_\pi^2)}{3},\\
B_\eta&=\frac{c_\theta(5M_K^2-3M_\pi^2)+4\sqrt{2}s_\theta M_K^2}{2\sqrt{6}},\\
B_{\eta'}&=\frac{s_\theta(5M_K^2-3M_\pi^2)-4\sqrt{2}c_\theta M_K^2}{2\sqrt{6}}.
\end{aligned}
\end{equation}
with $c_\theta=\cos\theta$ and $s_\theta=\sin\theta$.

\begin{table}[H]
\centering
\caption{Coefficients in the $(S,I)=(0,1/2)$ unitarized coupled-channel amplitude, taken from Ref.~\cite{Guo:2015dha}. }
\label{tab:potential-coefficients}
\small
\setlength{\tabcolsep}{5pt}
\renewcommand{\arraystretch}{1.25}
\begin{tabular}{lccccc}
\hline\hline
Channel & $C_{\rm LO}$ & $C_0$ & $C_1$ & $C_{24}$ & $C_{35}$ \\
\hline
$D\pi\to D\pi$ & $-2$ & $M_\pi^2$ & $-M_\pi^2$ & $1$ & $1$ \\
$D\eta\to D\eta$ & $0$ & $A_\eta$ & $-M_\pi^2X_\eta^2/3$ & $1$ & $X_\eta^2/3$ \\
$D_s\bar K\to D_s\bar K$ & $-1$ & $M_K^2$ & $-M_K^2$ & $1$ & $1$ \\
$D\eta'\to D\eta'$ & $0$ & $A_{\eta'}$ & $-M_\pi^2X_{\eta'}^2/3$ & $1$ & $X_{\eta'}^2/3$ \\
\hline
$D\eta\to D\pi$ & $0$ & $0$ & $M_\pi^2X_\eta$ & $0$ & $-X_\eta$ \\
$D_s\bar K\to D\pi$ & $-\sqrt{6}/2$ & $0$ & $-\sqrt{6}(M_K^2+M_\pi^2)/4$ & $0$ & $\sqrt{6}/2$ \\
$D_s\bar K\to D\eta$ & $-\sqrt{6}c_\theta/2$ & $0$ & $B_\eta$ & $0$ & $-(2\sqrt{2}s_\theta+c_\theta)/\sqrt{6}$ \\
$D\eta'\to D\pi$ & $0$ & $0$ & $-M_\pi^2X_{\eta'}$ & $0$ & $X_{\eta'}$ \\
$D\eta\to D\eta'$ & $0$ & $2(M_\pi^2-M_K^2)Z_{\eta\eta'}/3$ & $-M_\pi^2Z_{\eta\eta'}/3$ & $0$ & $Z_{\eta\eta'}/3$ \\
$D_s\bar K\to D\eta'$ & $-\sqrt{6}s_\theta/2$ & $0$ & $B_{\eta'}$ & $0$ & $(2\sqrt{2}c_\theta-s_\theta)/\sqrt{6}$ \\
\hline\hline
\end{tabular}
\end{table}

With the above potential, we search for poles in the unitarized coupled-channel amplitude and extract the scattering lengths. Following Ref.~\cite{Geng:2010vw}, 
 the scattering length of channel $i$ at its threshold is given  by
\begin{equation}
a_i
=-\frac{T_{ii}(s_{\mathrm{th},i})}{8\pi\sqrt{s_{\mathrm{th},i}}},
\qquad
\sqrt{s_{\mathrm{th},i}}=M_i+m_i .
\end{equation}
Here $M_i$ and $m_i$ are the masses of the charmed meson and the light pseudoscalar meson in channel $i$, respectively. 
Finally, we determine the couplings between the molecular states and their constituents from the pole residues,
\begin{eqnarray}
\label{couplingsdefine}
g_{i}g_{j}=\lim_{s\to s_0}\left(s-s_0\right)T_{ij}(s),
\end{eqnarray}
where $g_{i}$ denotes the coupling of channel $i$ to the dynamically generated state and $\sqrt{s_0}$ is the pole position. 
The obtained residues $g_i$ are  inputs for calculating the decay constants in Sec.~\ref{sec:decayconstants}.

\subsection{Amplitudes for  Cabibbo-favored  decays}\label{sec:weakapplications}

After determining the decay constants of the two-pole $D_0^*(2300)$, we proceed with their phenomenological applications to $b$-flavored  hadron   decays. In this work,  we concentrate exclusively on Cabibbo-favored decays, for which nonfactorizable corrections are expected to be comparatively small~\cite{Yuan:2025pnt}.

\begin{figure}[H]
\centering
\subfigure[]{
\includegraphics[width=0.39\textwidth]{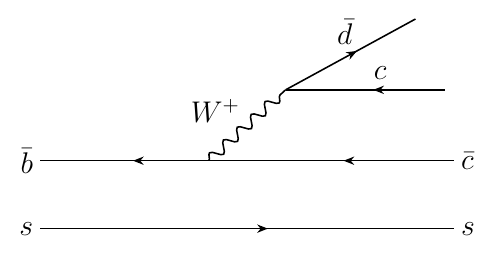}
}
\subfigure[]{
\includegraphics[width=0.39\textwidth]{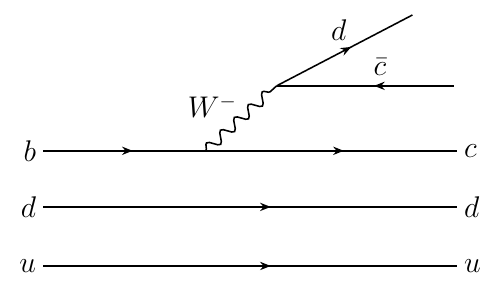}
}
\caption{External $W$-emission mechanism for 
    (a) $B_s^0\to D_s^-D_0^{*+}$ and 
    (b) $\Lambda_b^0\to\Lambda_c^+D_0^{*-}$. 
    The corresponding $\Xi_b^{0,-}\to\Xi_c^{+,0}D_0^{*-}$ processes
    are obtained from panel (b) by replacing the spectator $d(u)$ quark with an $s$ quark.}
\label{Fig:Tree}
\end{figure}


At the quark level, the Cabibbo-favored decays
$B_s^0\to D_s^-D_0^{*+}$, 
$\Lambda_b^0\to\Lambda_c^+D_0^{*-}$, and
$\Xi_b^{0,-}\to\Xi_c^{+,0}D_0^{*-}$
proceed mainly through the external $W$-emission mechanism shown in Fig.~\ref{Fig:Tree}.
According to the topological classification of weak decays, these
processes can be described within the naive factorization
approach~\cite{Bauer:1986bm,Cheng:1996cs,Beneke:2003zv}.
In the factorization ansatz, the amplitudes of the weak decays 
$B_s^0\to D_s^-D_0^{*+}$ and 
$\Lambda_b^0\to\Lambda_c^+D_0^{*-}$ 
can be expressed as products of two current matrix elements,
\begin{align}\label{Ds-KK}
\mathcal{A}\left(B_s^0 \to D_s^- D_{0}^{*+}\right)&=\frac{G_{F}}{\sqrt{2}} V_{cb}V_{cd}^{*} a_{1}\left\langle D_{0}^{*+}|(\bar{c}d)| 0\right\rangle\left\langle D_s^{-}|(\bar{b}c)| B_s^{0}\right\rangle, \\
\mathcal{A}\left(\Lambda_b^0 \to \Lambda_c^+ D_{0}^{*-}\right)&=\frac{G_{F}}{\sqrt{2}} V_{cb}V_{cd}^{*} a_{1}\left\langle D_{0}^{*-}|(\bar{d}c)| 0\right\rangle\left\langle \Lambda_c^+|(\bar{c}b)| \Lambda_b^0\right\rangle, 
\end{align}
where $G_F$ is the Fermi constant, and $V_{cb}$ and $V_{cd}$ are the Cabibbo--Kobayashi--Maskawa matrix elements. Here $(\bar q_1 q_2)$ denotes the corresponding weak $V-A$ current.
In this work, we adopt the effective Wilson coefficient $a_1=1.03$~\cite{Yuan:2025pnt,Lu:2009cm}. The current matrix element creating $D_{0}^{*}$ from the vacuum is parametrized by the decay constant in Eq.~(\ref{eq:decayconstant}).

The current matrix element $\left\langle D_s|(\bar{b}c)| B_s\right\rangle$, which describes the hadronic transition, is parametrized by two form factors~\cite{Cheng:2006dm},
\begin{equation}
\left\langle D_s|(\bar{b}c)| B_s\right\rangle=\left[P^{\mu}-\frac{m_{B_s}^2-m_{D_s}^2}{q^2}q^{\mu}\right] F_{1}(q^2)+\frac{m_{B_s}^2-m_{D_s}^2}{q^2}q^{\mu} F_{0}(q^2),   
\end{equation}
where $q=p_{B_s^0}-p_{D_s^-}$ and $P=p_{B_s^0}+p_{D_s^-}$, and $F_{0}(q^2)$ and $F_{1}(q^2)$ are form factors.   
The current matrix element $\langle \Lambda_c(p^{\prime})|\bar{c} b|\Lambda_b(p)\rangle$ is given by~\cite{Gutsche:2018utw} 
\begin{equation}
\begin{aligned}
\langle \Lambda_c(p^{\prime})|(\bar{c}b)|\Lambda_b(p)\rangle
&=\bar{u}(p^{\prime})[f_{1}^V(q^2)\gamma^{\mu}
-f_2^V(q^2)\frac{i\sigma^{\mu\nu}q_{\nu}}{m}
+f_3^V(q^2)\frac{q^{\mu}}{m}    \\ 
&-(f_{1}^A(q^2)\gamma^{\mu}
-f_2^A(q^2)\frac{i\sigma^{\mu\nu}q_{\nu}}{m}
+f_3^A(q^2)\frac{q^{\mu}}{m})\gamma^5]u(p).
\end{aligned}
\end{equation}
where $\sigma^{\mu\nu}=\frac{i}{2}(\gamma^\mu\gamma^\nu-\gamma^\nu\gamma^\mu)$, $q=p-p^{\prime}$, and $f_{i}^{V/A}(q^2)$ are form factors. For the retained phenomenological $\Xi_b\to\Xi_c$ input, the form factors are parametrized as
\begin{equation}\label{formfactors}
F_{i}(q^2)=\frac{F_i(0)}{1-a\, \frac{q^2}{m^2}+ b\, (\frac{q^2}{m^2})^2}, 
\end{equation}
where $F_{i}(0)$, $a$, and $b$ are the phenomenological parameters listed in Table~\ref{BtoKformfactor1123}.

The amplitude for $B_s \to \bar{D}_s D_{0}^{*}$ is then
\begin{equation}\label{am3}
\mathcal{A}(B_s \to \bar{D}_s D_{0}^{*})=\frac{G_{F}}{\sqrt{2}}V_{cb}V_{cd}^{*} a_{1}f_{D_{0}^{*}}(m_{B_s}^2-m_{\bar{D}_s}^2) F_{0}(q^2).
\end{equation}

For $\Lambda_b( p)\rightarrow\Lambda_c(p^{\prime}) D_{0}^{*}(q)$, the corresponding amplitude can be written as~\cite{Cheng:1996cs}
\begin{equation}\label{Eq:weakdecayV1}
\mathcal{A}(\Lambda_b \to \Lambda_cD_{0}^{*})=\bar{u}(p^{\prime})(A+B \gamma_5)u(p),
\end{equation}
where $A$ and $B$ are expressed in terms of the $\Lambda_b\to\Lambda_c$ transition form factors as
\begin{equation}\label{Eq:weakdecayV2}
\begin{split}
    A=\lambda f_{D_{0}^{*}}[(m-m_2)f_1^V+\frac{m_1^2}{m}f_3^V], \\
    B=\lambda f_{D_{0}^{*}}[(m+m_2)f_1^A-\frac{m_1^2}{m}f_3^A],
\end{split}
\end{equation}
with $\lambda=\frac{G_{F}}{\sqrt{2}}V_{cb}V_{cd}^{*} a_{1}$.  Here  $m$, $m_1$, and $m_2$ denote the masses of $\Lambda_b$, $D_{0}^{*}$, and $\Lambda_c$, respectively.  
Assuming SU(3)-flavor symmetry, the $\Lambda_b \to \Lambda_c$ transition can be related to the $\Xi_b \to \Xi_c$ transition~\cite{Geng:2018plk}. Since the production mechanism of the $D_{0}^{*}$ in $\Xi_b$ decay is analogous to that in $\Lambda_b$ decay, the corresponding decay amplitudes share the same form.

 The corresponding partial decay widths can be obtained from
 \begin{eqnarray}
\Gamma=\frac{1}{2J+1}\frac{1}{8\pi}\frac{|\vec{p}|}{{ M}^2}{\overline{|{\mathcal{A}}|}}^{2},
\end{eqnarray}
where $J$ and $M$ are the total angular momentum and mass of the initial state, $\vec{p}$ is the three-momentum of either final particle in the rest frame of the initial state, and the overline denotes the spin sum over final states.     


\section{Results and discussion}\label{III}
\label{secresults}

\begin{table}[H]
\centering
\caption{ Values of unknown parameters in unitarized coupled-channel  amplitude.}
\label{tab:potential-inputs}
\renewcommand{\arraystretch}{1.25}
\begin{tabular}{cccc}
\hline\hline
Parameter & Value & Parameter & Value \\
\hline
$F$ & $92.2~\mathrm{MeV}$ & $\mu$ & $1.0~\mathrm{GeV}$ \\
$\bar M_D$ & $\simeq1.92~\mathrm{GeV}$ & $\theta$ & $\simeq-19.0^\circ$ \\
$h_0$ & $0.033$ & $h_1$ & $0.43$ \\
$h_2$ & $0.08$ & $h_3$ & $3.79$ \\
$h_4$ & $\simeq-0.057~\mathrm{GeV}^{-2}$ & $h_5$ & $\simeq-0.484~\mathrm{GeV}^{-2}$ \\
$h_{24}$ & $-0.13^{+0.05}_{-0.06}$ & $h_{35}$ & $0.23\pm0.06$ \\
$\hat h_4$ & $-0.21^{+0.29}_{-0.27}$ & $\hat h_5$ & $-1.78\pm0.19$ \\
$a(1~\mathrm{GeV})$ & $-1.88\pm0.06$ & & \\
\hline\hline
\end{tabular}
\end{table}

\begin{figure}[H]
\centering
\includegraphics[width=0.58\textwidth]{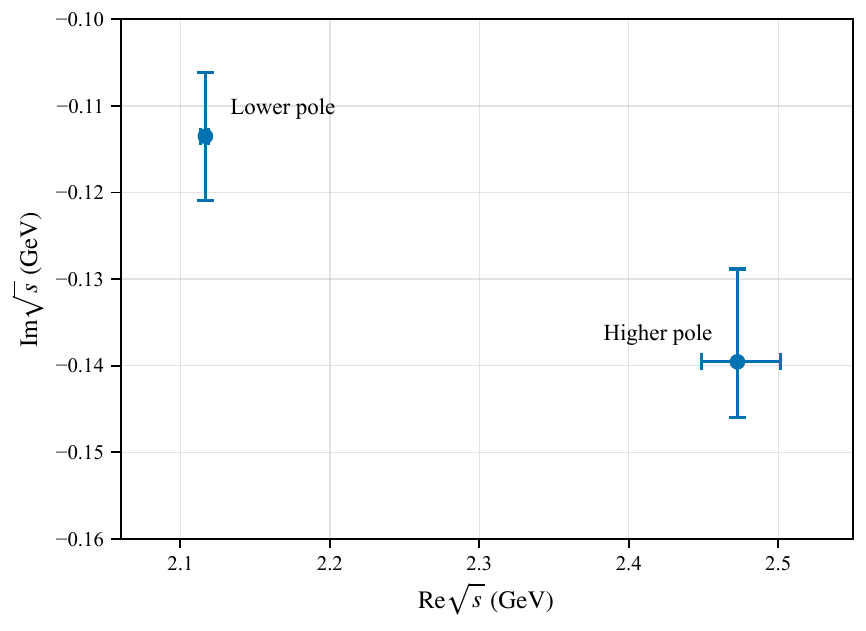}
\caption{Pole locations of the two dynamically generated states in the
complex $\sqrt{s}$ plane. The horizontal and vertical error bars denote
the asymmetric uncertainties of $\mathrm{Re}\sqrt{s}$ and
$\mathrm{Im}\sqrt{s}$, respectively.}
\label{fig:polesreply}
\end{figure}

 At first, we   study the two-pole $D_0^*(2300)$ generated by the $D\pi$, $D\eta$, $D_s\bar K$, and $D\eta'$ coupled-channel amplitude.
Following the Fit-6C parametrization of Ref.~\cite{Guo:2015dha}, we adopt the values of the relevant  parameters listed in Table~\ref{tab:potential-inputs}\footnote{{ Rewriting the Feynman-parameter representation of Eq.~\eqref{loopfunctionv} in the
analytic dimensional-regularization scheme of Ref.~\cite{Guo:2015dha} shows
that it corresponds to the implicit subtraction choice $a(\mu)=-2$.
The Fit-6C prescription is therefore implemented as   $G(s;\mu)+\frac{a(\mu)+2}{16\pi^2}$
with $\mu=1~\mathrm{GeV}$ and $a(1~\mathrm{GeV})=-1.88\pm0.06$ from
Ref.~\cite{Guo:2015dha}. The full loop function is independent of the
renormalization scale: the explicit $\mu$ dependence is compensated by
the running of the subtraction constant, $a(\mu')=a(\mu)+\ln\frac{\mu'^2}{\mu^2}$, 
so that $G(s;\mu')-G(s;\mu)=0$. We thus fix $\mu=1~\mathrm{GeV}$ as in
Fit-6C and propagate the fitted uncertainty of $a(\mu)$ together with
its correlations with the remaining Fit-6C parameters. }}. The uncertainties in the pole positions arise from the correlated errors in the five Fit-6C parameters $h_{24}$, $h_{35}$, $\hat h_4$, $\hat h_5$, and $a(\mu)$\footnote{Asymmetric input uncertainties are symmetrized, and their correlations are propagated using multivariate Gaussian sampling based on the five-parameter Fit-6C correlation submatrix. The quoted errors include these five correlated parameter uncertainties. }.
As shown in Table~\ref{moleculecouplingsc}, the lower pole is located at \(\sqrt{s^{(\mathrm{lower})}}=(2116.69^{+2.36}_{-3.18})-i(113.51^{+7.39}_{-7.37})~\mathrm{MeV}\), while the higher pole appears at \(\sqrt{s^{(\mathrm{higher})}}=(2472.68^{+28.99}_{-24.25})-i(139.55^{+6.43}_{-10.72})~\mathrm{MeV}\). The two-pole structure is naturally generated in our calculations as shown in Fig.~\ref{fig:polesreply}, and the pole positions are consistent with those in Refs.~\cite{Albaladejo:2016lbb,Du:2017zvv,Zhuang:2026lta}. With the above amplitude, we further calculate the \(I=1/2\) \(D\pi\) scattering length, \(a_{D\pi}=0.3455^{+0.0098}_{-0.0087}~\mathrm{fm}\), consistent with Refs.~\cite{Flynn:2007ki,Guo:2009ct,Geng:2010vw,Moir:2016srx}\footnote{By simulating the Lattice QCD data and using the ChEFT potentials, the \(I=1/2\) \(D\pi\) scattering length is determined as \(a_{D\pi}=0.3~\mathrm{fm}\) without unitarized amplitude~\cite{Geng:2010vw}, \(a_{D\pi}=0.37\pm0.01~\mathrm{fm}\)~\cite{Liu:2012zya}, and \(a_{D\pi}=0.36(1)~\mathrm{fm}\)~\cite{Guo:2009ct} with the unitarized amplitude.}, which demonstrates that the obtained scattering amplitude is reliable. Finally, we calculate the two-pole couplings to their constituents, as shown in Table~\ref{moleculecouplingsc}. One can see that the lower pole couples most strongly to $D\pi$, whereas the higher pole is dominated by $D_s\bar{K}$, consistent with Refs.~\cite{Guo:2015dha,Albaladejo:2016lbb}.

\begin{table}[H]
\centering
\caption{Pole positions (in MeV) and couplings to constituents (in GeV) for the two-pole $D_0^*(2300)$.  }
\label{moleculecouplingsc}
\small
\renewcommand{\arraystretch}{1.25}
\begin{tabular}{lcccccc}
\hline\hline
Pole & $M$ (MeV) & $\Gamma/2$ (MeV) & $\lvert g_{D\pi}\rvert$ & $\lvert g_{D\eta}\rvert$ & $\lvert g_{D_s\bar K}\rvert$ & $\lvert g_{D\eta'}\rvert$ \\
\hline
Lower  & $2116.69^{+2.36}_{-3.18}$ & $113.51^{+7.39}_{-7.37}$ & $9.724^{+0.129}_{-0.156}$ & $3.147^{+0.217}_{-0.275}$ & $4.448^{+0.283}_{-0.293}$ & $5.985^{+0.531}_{-0.629}$ \\
Higher & $2472.68^{+28.99}_{-24.25}$ & $139.55^{+6.43}_{-10.72}$ & $5.355^{+0.403}_{-0.293}$ & $5.800^{+0.660}_{-0.405}$ & $11.339^{+0.606}_{-0.445}$ & $6.259^{+0.613}_{-0.578}$ \\
\hline\hline
\end{tabular}
\end{table}

After determining the pole positions, their couplings to the constituent mesons, and the $D_{(s)}P$ weak-transition form factors extracted from semileptonic data, we evaluate the decay constants of the two-pole $D_0^*(2300)$ using Eq.~\eqref{fd2317coupling}. As shown in the analysis of Ref.~\cite{Liu:2023cwk}, the decay constant of the two-pole $D_0^*(2300)$ is independent of the renormalization scale $\mu$\footnote{ The logarithm in the decay-constant expression of Eq.~\eqref{fd2317coupling} is split as
$\ln[\Delta^2(x;p^2)/\mu^2]=\ln\Delta^2(x;p^2)-\ln\mu^2$. The first
term is independent of the renormalization scale $\mu$, whereas the
second term depends on $\mu$ explicitly. However, the latter is
multiplied by $\int_0^1dx\,(2x-1)=0$ and thus vanishes identically.
Therefore, the decay constant is independent of $\mu$.  }. We therefore adopt the same renormalization scale as that used for the loop function in the scattering equation~\cite{Guo:2015dha}. For the lower pole, we obtain the decay constant $f_{D_0^*}^{(\mathrm{lower})}=68.46^{+0.93}_{-1.25}\,\mathrm{MeV}$, while for the higher pole we find $f_{D_0^*}^{(\mathrm{higher})}=90.51^{+5.91}_{-11.64}\,\mathrm{MeV}$. These errors include the correlated five-parameter Fit-6C uncertainty from $h_{24}$, $h_{35}$, $\hat h_4$, $\hat h_5$, and $a(\mu)$, together with the semileptonic-branching-fraction-induced uncertainty of the local weak inputs. As indicated in Ref.~\cite{Albaladejo:2016lbb}, the lower pole and the $DK$ bound state belong to the same $\bar{\mathbf 3}$ representation. We have estimated the decay constant of the $DK$ molecule to be $59$~MeV~\cite{Liu:2023cwk}. Its numerical proximity to the lower-pole decay constant is qualitatively consistent with the expected SU(3)-flavor relation.

  \begin{table}[H]
 \centering
 \caption{ Values of the $D_{0}^*(2300)$ decay constants  from  different approaches (in units of MeV). \label{Tab:Dcs} }
 \renewcommand{\arraystretch}{1.3}
 \begin{tabular}{cc}
 \hline\hline
 Approaches ~~~  & Decay Constants   \\
 \hline
 QCD sum rule~\cite{Wang:2015mxa}~~~  &$373\pm 19$~~~  \\
 Quark model~\cite{Veseli:1996yg}~~~  &$139\pm 30$~~~  \\
Salpeter method~\cite{Wang:2007av}~~~   &$133$~~~   \\ 
Covariant light-front quark model~\cite{Verma:2011yw}~~~   &$107\pm 13$~~~    \\ 
Ours~~~   &$68.46^{+0.93}_{-1.25}/90.51^{+5.91}_{-11.64}$~~~  \\ 
\hline\hline
 \end{tabular}
 \end{table}

In this work, we determine the decay constants of the two-pole $D_0^*(2300)$ in the molecular picture. As shown in Table~\ref{Tab:Dcs}, the decay constant of $D_0^*(2300)$ has been estimated by the QCD sum rule~\cite{Wang:2015mxa}, the quark model~\cite{Veseli:1996yg}, the Salpeter method~\cite{Wang:2007av}, and the covariant light-front quark model~\cite{Verma:2011yw}, where $D_0^*(2300)$ is treated as a conventional excited charmed meson. We find that the decay constant of $D_0^*(2300)$ when treated as a hadronic molecule is smaller than that when treated as an excited charmed meson, indicating that the decay constant is sensitive to the internal structure of $D_0^*(2300)$. 
 Furthermore, we examine the SU(3)-flavor symmetry breaking of the decay
constants, whose pattern reflects the irrep content of the two poles. For the
higher pole, the matrix element of the local current $(c \bar d)$ vanishes in
the SU(3)-flavor-symmetric limit~\cite{Albaladejo:2016lbb,Zhuang:2026lta}; its physical value is therefore
entirely generated by explicit breaking, and
$f_{D_0^*}^{(\mathrm{higher})}$ measures the $\bar{\mathbf 3}$ admixture of
the higher-pole wave function. This admixture is naturally large: the higher
pole couples to the strange-sector channels $D\eta$, $D_s\bar K$ and
$D\eta'$, where the mass splittings ($m_K-m_\pi\simeq354$~MeV,
$m_\eta-m_\pi\simeq408$~MeV and $M_{D_s}-M_D\simeq104$~MeV) far exceed the
weak attraction of the $\mathbf 6$ sector~\cite{Zhuang:2026lta}. In contrast, within the
$\bar{\mathbf 3}$ multiplet the breaking is moderate: our
$f_{D_0^*}^{(\mathrm{lower})}=68.46$~MeV and the $DK$-molecule decay constant
$f_{DK}\simeq59$~MeV of Ref.~\cite{Liu:2023cwk}, the $\bar{\mathbf 3}$ partner
of the lower pole, differ by only $\sim15\%$, in line with the almost purely
molecular character of the $\bar{\mathbf 3}$ pole found in the SU(3) region of
Ref.~\cite{Zhuang:2026lta}. For comparison, in the approaches where the
$D_0^*(2300)$ is treated as a conventional excited charmed meson --- the  QCD sum rule~\cite{Wang:2015mxa}, quark model~\cite{Veseli:1996yg}, Salpeter method~\cite{Wang:2007av}, and covariant light-front quark model~\cite{Verma:2011yw} --- the corresponding SU(3) breaking amounts to
$11\%$, $26\%$, $19\%$, and $45\%$, respectively.

 To quantify the flavor composition of the two physical poles, we define
effective mixing angles from their complex residue couplings in the full
four-channel space. For each pole $p=L,H$, denoting the lower and higher
poles, respectively, the residue couplings are extracted as in Eq.~\eqref{couplingsdefine},
\begin{equation}
T_{ij}(s)=\frac{g_i^{(p)}g_j^{(p)}}{s-s_p}+T_{ij}^{\mathrm{reg}}(s),
\qquad
\boldsymbol g^{(p)}=\big(g_{D\pi}^{(p)},\,g_{D\eta}^{(p)},\,
g_{D_s\bar K}^{(p)},\,g_{D\eta'}^{(p)}\big)^{T} .
\label{eq:residue-couplings}
\end{equation}
All transformations below act on the full complex couplings in the
normalized channel basis, with a common phase convention and without
taking individual coupling magnitudes before projection. Suppressing the
pole label, we first rotate the $\eta,\eta'$ channels to the octet and
singlet basis,
\begin{equation}
g_8=c_\eta\, g_{D\eta}+s_\eta\, g_{D\eta'},\qquad
g_1=-s_\eta\, g_{D\eta}+c_\eta\, g_{D\eta'},
\label{eq:eta-rot}
\end{equation}
where $(c_\eta,s_\eta)\simeq(0.9456,-0.3254)$ are the cosine and sine of
the fixed $\eta$--$\eta'$ mixing angle used in the interaction kernel
(cf. Table~\ref{tab:potential-inputs}). The SU(3) projections are then \cite{Zhuang:2026lta}
\begin{equation}
\begin{pmatrix}a_{\bar3,8}\\ a_6\\ a_{\overline{15}}\end{pmatrix}
=
\begin{pmatrix}
-3/4 & -1/4 & -\sqrt{3/8}\\
\sqrt{3/8} & -\sqrt{3/8} & -1/2\\
1/4 & 3/4 & -\sqrt{3/8}
\end{pmatrix}
\begin{pmatrix}g_{D\pi}\\ g_8\\ g_{D_s\bar K}\end{pmatrix},
\qquad a_{\bar3,1}=g_1 .
\label{eq:su3-projection}
\end{equation}
The full flavor space contains two antitriplet copies,
\begin{equation}
\overline{\mathbf3}\otimes(\mathbf8\oplus\mathbf1)
=\overline{\mathbf3}_{8}\oplus\mathbf6
 \oplus\overline{\mathbf{15}}\oplus\overline{\mathbf3}_{1},
\label{eq:decomposition}
\end{equation}
and the singlet component $a_{\bar3,1}$, originating from the $D\eta_0$
channel, is retained in the total antitriplet strength. We define the
total antitriplet norm and the effective mixing angle of pole $p$ by
\begin{equation}
A_{\bar3,p}=
\sqrt{|a_{\bar3,8}^{(p)}|^2+|a_{\bar3,1}^{(p)}|^2},
\qquad
\theta_p=\operatorname{atan2}\!\left(|a_6^{(p)}|,\,A_{\bar3,p}\right),
\qquad p=L,H .
\label{eq:mixing-angle}
\end{equation}
Here $\theta_p=0^\circ$ denotes the antitriplet direction and
$\theta_p=90^\circ$ the sextet direction within the combined
antitriplet--sextet subspace. These angles are invariant under a common
nonzero complex rescaling of the residue vector and under unitary
rotations between the two antitriplet copies. They summarize relative
residue-coupling strengths, not a complete complex mixing transformation
or wave-function probabilities: the relative complex phases enter the
projections in Eq.~\eqref{eq:su3-projection} but are not encoded by the
angles themselves. For completeness we also report the normalized
residue-coupling weights
\begin{equation}
w_{\alpha,p}=\frac{|a_{\alpha}^{(p)}|^2}{N_p},\qquad
N_p=|a_{\bar3,8}^{(p)}|^2+|a_6^{(p)}|^2+
|a_{\overline{15}}^{(p)}|^2+|a_{\bar3,1}^{(p)}|^2 ,
\label{eq:weights}
\end{equation}
in which the $\overline{\mathbf{15}}$ component is included, while it is
excluded from the two-subspace angle in Eq.~\eqref{eq:mixing-angle}.

\begin{table}[t]
\caption{\label{tab:mixing}%
Effective antitriplet--sextet mixing angles $\theta_p$ and normalized
residue-coupling weights $w_{\alpha,p}$ of the two poles, evaluated at
the central Fit-6C parameters. The angle uncertainties are obtained by
propagating the correlated uncertainties of the five Fit-6C parameters
$(h_{24},h_{35},\hat h_4,\hat h_5,a)$ using multivariate Gaussian
sampling (see text). The weights are central values and sum to unity up
to rounding.}
\begin{ruledtabular}
\begin{tabular}{lccccc}
Pole & $\theta_p$ (deg) & $w_{\bar3,8}$ & $w_6$ &
$w_{\overline{15}}$ & $w_{\bar3,1}$\\
\colrule
Lower  & $20.09^{+0.91}_{-0.91}$ & 0.5961 & 0.1169 & 0.0089 & 0.2781\\
Higher & $36.78^{+0.75}_{-0.44}$ & 0.5229 & 0.3451 & 0.0371 & 0.0948\\
\end{tabular}
\end{ruledtabular}
\end{table}

The numerical results are collected in Table~\ref{tab:mixing}. The
uncertainties of the effective angles are obtained by propagating the
correlated uncertainties of the five Fit-6C parameters
$(h_{24},h_{35},\hat h_4,\hat h_5,a)$ through multivariate Gaussian
sampling, with asymmetric input uncertainties symmetrized and the
subtraction constant and its correlations included, while the hadron
masses and the $\eta$--$\eta'$ mixing input are held fixed. The quoted
errors correspond to the $68\%$ central percentile intervals of the
resulting angle distributions and include only these correlated
fit-parameter uncertainties. The relation between the angle and the
weights reads
\begin{equation}
\sin^2\theta_p=\frac{w_{6,p}}{1-w_{\overline{15},p}},\qquad
\cos^2\theta_p=\frac{w_{\bar3,8,p}+w_{\bar3,1,p}}
{1-w_{\overline{15},p}} .
\label{eq:angle-weight}
\end{equation}
Thus $\sin^2\theta_p$ is the sextet fraction of the squared
residue-coupling norm within the selected antitriplet--sextet
subspace, rather than its weight in the complete four-component
decomposition. We emphasize that the residue couplings are complex;
evaluating Eq.~\eqref{eq:su3-projection} with only their magnitudes
from Table~\ref{moleculecouplingsc} would yield appreciably different values, since the
relative phases of the couplings enter the projections.

For the lower pole, the combined antitriplet and sextet weights are
approximately $0.874$ and $0.117$, respectively, while for the higher
pole they are approximately $0.618$ and $0.345$; the corresponding
$\overline{\mathbf{15}}$ weights are $0.0089$ and $0.0371$. The lower
pole is therefore predominantly antitriplet in this residue-based sense,
whereas the higher pole has a considerably larger sextet contribution
but still retains substantial antitriplet strength, and is not sextet
dominated at the physical point according to this definition. The two
effective angles thus quantify a pronounced difference between the
physical-pole residue compositions. In particular, the sizeable
antitriplet fraction of the higher pole, $\cos^2\theta_H\simeq0.64$, is
consistent with the large $\bar3$ admixture measured by
$f_{D_0^*}^{(\mathrm{higher})}$ discussed above.

This physical-point decomposition is distinct from the assignment of a
pole's dynamical origin through continuation to the SU(3) limit, and
does not by itself determine the poles' SU(3)-limit assignments. In
exact SU(3), the antitriplet weak current cannot couple to a pure sextet
state; if a pole connects to the sextet sector in the SU(3) limit,
symmetry breaking can generate antitriplet components that contribute to
its physical weak-current coupling. The results in Table~\ref{tab:mixing}
quantify such residue mixing, whereas the weak decay constant involves
the coherent weak-source projection, including relative phases and
symmetry-breaking terms, so that its magnitude is not determined by the
residue angle alone.



\begin{table}[H]
\centering
\caption{Branching fractions ($10^{-4}$) of the decays $B_s \to D_s D_0^*(2300)$, $\Lambda_b \to \Lambda_c D_0^*(2300)$, and $\Xi_b \to \Xi_c D_0^*(2300)$.  
\label{resultsbsto2317}}
\renewcommand{\arraystretch}{1.3}
\begin{tabular}{c c | c c}
  \hline \hline
  Decay modes & Exp~\cite{ParticleDataGroup:2026mpi} & Decay modes & Ours \\
  \hline
  \multirow{2}{*}{$B_s^0 \to D_s^- D^+$}
  & \multirow{2}{*}{$3.1\pm0.5$}
  & $B_s^0 \to D_s^- D_{0}^{*+(\mathrm{lower})}$ & $0.534^{+0.021}_{-0.025}$ \\
  & 
  & $B_s^0 \to D_s^- D_{0}^{*+(\mathrm{higher})}$ & $0.880^{+0.115}_{-0.208}$ \\
  \hline

  Decay modes & Exp~\cite{ParticleDataGroup:2026mpi} & Decay modes & Ours \\
  \hline
  \multirow{2}{*}{$\Lambda_b^0 \to \Lambda_c^+D^-$}
  & \multirow{2}{*}{$4.6\pm0.6$}
  & $\Lambda_b^0 \to \Lambda_c^+D_{0}^{*-(\mathrm{lower})}$ & $0.499^{+0.044}_{-0.048}$ \\
  &
  & $\Lambda_b^0 \to \Lambda_c^+D_{0}^{*-(\mathrm{higher})}$ & $0.865^{+0.129}_{-0.202}$ \\
  \hline

  Decay modes & Ours & Decay modes & Ours \\
  \hline
  \multirow{2}{*}{$\Xi_b^{0,-} \to \Xi_c^{+,0}D^-$}
  & \multirow{2}{*}{$5.7$}
  & $\Xi_b^{0,-} \to \Xi_c^{+,0}D_{0}^{*-(\mathrm{lower})}$ & $0.578^{+0.016}_{-0.021}$ \\
  &
  & $\Xi_b^{0,-} \to \Xi_c^{+,0}D_{0}^{*-(\mathrm{higher})}$ & $0.979^{+0.126}_{-0.233}$ \\
  \hline \hline
\end{tabular}
\end{table}

We further calculate the branching fractions of Cabibbo-favored decays of $b$-flavored hadrons into $D_0^*(2300)$, which provide a test of the calculated decay constants and may help discriminate between different interpretations of its internal structure. 
For the $B_s\to D_s$ transition, we directly use the HPQCD lattice-QCD form factor of Ref.~\cite{McLean:2019qcx}. At the two central $D_0^*$ pole masses, $f_0(M_{\rm lower}^2)=0.746$ and $f_0(M_{\rm higher}^2)=0.780$, with the published covariance propagated. For $\Lambda_b\to\Lambda_c$, we directly use the lattice-QCD helicity form factors of Ref.~\cite{Detmold:2015aaa}. At the two central pole masses, $(f_0,g_0)_{\rm lower}=(0.597,0.575)$ and $(f_0,g_0)_{\rm higher}=(0.677,0.665)$, with the published statistical covariance and systematic uncertainty propagated. For $\Xi_b\to\Xi_c$, we retain the phenomenological transition form factors, whose parameters are listed in Table~\ref{BtoKformfactor1123}. The quoted $\Xi_b$ uncertainties do not include an additional transition-form-factor model error.

Our calculated branching fractions are listed in Table~\ref{resultsbsto2317}. For the lower and higher poles in the $B_s$ decay, we obtain $0.534\times10^{-4}$ and $0.880\times10^{-4}$, respectively. For the lower pole, the branching fractions in the $\Lambda_b$ and $\Xi_b$ decays are $0.499\times10^{-4}$ and $0.578\times10^{-4}$, respectively, which are similar to that in the $B_s$ case. For the higher pole, the branching fractions in the $\Lambda_b$, $\Xi_b$, and $B_s$ decays are $0.865\times10^{-4}$, $0.979\times10^{-4}$, and $0.880\times10^{-4}$, respectively, and are all larger than those of the lower pole. In comparison, the branching fractions for the corresponding ground-state charmed mesons are roughly one order of magnitude larger. These similarities suggest that the branching fractions are primarily governed by the decay constants of the produced charmed states.    The two
poles possess similar production mechanisms in $b$-flavored hadron
decays, so that the two-pole structure should be visible in the
invariant mass distribution.   
If the decay constant of the higher
pole were strongly suppressed, this pole would not manifest itself in
the invariant mass distribution.  Analysis of the invariant mass
distribution suggests the presence of a two-pole structure~\cite{Du:2019oki,Lyu:2023ppb}, which indicate it's likely to confirm the existence of two-pole structure in these decays in terms of our results.


\begin{table}[ttt]
\centering

 \caption{Values of the phenomenological parameters $F(0)$, $a$, and $b$ for the $\Xi_b\rightarrow\Xi_c$ transition form factors~\cite{Faustov:2018ahb}. \label{BtoKformfactor1123}}
 \renewcommand{\arraystretch}{1.3}
 \setlength{\tabcolsep}{12pt}
 \begin{tabular}{c|cccc}
 \hline\hline
   & $f_1^V$ & $f_3^V$ & $f_1^A$ & $f_3^A$ \\
 \hline
 $F(0)^{\Xi_b\to\Xi_c}$ & 0.467 & 0.086 & 0.447 & -0.278 \\
 $a^{\Xi_b\to\Xi_c}$ & 1.702 & 1.742 & 1.759 & 2.270 \\
 $b^{\Xi_b\to\Xi_c}$ & 0.531 & 0.758 & 0.356 & 1.072 \\
\hline\hline
 \end{tabular}

 \end{table}



\section{Summary}\label{IV}

The excited charmed meson $D_0^*(2300)$ poses a notable challenge to the conventional quark model and simultaneously provides a unique window into the nonperturbative binding dynamics of the strong interaction.  In this work, we computed its decay constants using the effective Lagrangian approach under the two-pole hypothesis. This two-pole structure is dynamically generated by the unitarized amplitudes of the coupled channels $D\pi$, $D\eta$, $D_s\bar K$, and $D\eta'$. The lower pole, at $\sqrt{s^{\rm (lower)}}=(2116.69^{+2.36}_{-3.18})-i(113.51^{+7.39}_{-7.37})~\mathrm{MeV}$, couples predominantly to the $D\pi$ channel, whereas the higher pole, at $\sqrt{s^{\rm (higher)}}=(2472.68^{+28.99}_{-24.25})-i(139.55^{+6.43}_{-10.72})~\mathrm{MeV}$, is dominated by the $D_s\bar K$ channel. Utilizing these pole properties, we determined their decay constants to be $f^{({\rm lower})}_{D_0^*}=68.46^{+0.93}_{-1.25}\ {\rm MeV}$ and $f^{({\rm higher})}_{D_0^*}=90.51^{+5.91}_{-11.64}\ {\rm MeV}$. These values are substantially smaller than the predictions from conventional $c\bar q$ excited-state models, suggesting that the decay constant can serve as a  discriminator between molecular and compact interpretations of the $D_0^*(2300)$. We further characterized the flavor composition of the two poles through
effective residue-based mixing angles,
$\theta_L=20.09^{+0.91}_{-0.91}\,\mathrm{deg}$ and
$\theta_H=36.78^{+0.75}_{-0.44}\,\mathrm{deg}$, measured from the
combined antitriplet direction. The lower pole is predominantly
antitriplet, while the higher pole carries a sizeable sextet
contribution yet retains the substantial antitriplet strength that
generates its physical weak decay constant. 

From a phenomenological perspective, we estimated the production rates of the two-pole $D_0^*(2300)$ in the Cabibbo-favored decays of the $B_s$, $\Lambda_b$, and $\Xi_b$ hadrons,  for which the factorization approach provides a reliable description. The resulting rates for these $b$-flavored hadron decays are found to be on the order of several $10^{-5}$, roughly one order of magnitude smaller than those of the corresponding ground-state charmed mesons, in agreement with the production ratio between ground and excited charmed-strange mesons in similar decays. Owing to its larger decay constant, the higher pole exhibits a correspondingly enhanced production rate. Future experimental measurements of these decays, especially the ratio  between the two poles and the ground-state charmed mesons, would offer a crucial test of the two-pole molecular scenario for $D_0^*(2300)$.

\section*{Acknowledgments}

This work is supported by the National Natural Science Foundation of China under Grant No.~12622505,  No.~12575086 and No. W2543006. Qi-Wei Yuan acknowledges support from  the
Fundamental Research Funds for the Central Universities under No.~lzujbky-2023-stlt01.

\appendix

\bibliography{reference}

\end{document}